\documentclass[10pt,twocolumn]{article}
\usepackage[margin=0.8in]{geometry}
\usepackage{amsmath,amssymb}
\usepackage{booktabs}
\usepackage{graphicx}
\usepackage{xcolor}
\usepackage[font=small,labelfont=bf]{caption}
\usepackage[colorlinks=true,citecolor=blue!50!black,linkcolor=blue!50!black,urlcolor=blue!50!black]{hyperref}
\usepackage{microtype}

\title{Tree Speculative Decoding with In-Call KV Repair\\on a Commercial Static-Graph NPU}
\author{Sungwon Woo \quad Seyeon Kim\thanks{Korea University. Corresponding author: Seyeon Kim.}}
\date{}

\begin{document}
\maketitle

\begin{abstract}
Tree speculative decoding verifies many draft branches in one target call, but common implementations assume dynamic query shapes and direct control over the KV cache. Static-graph NPUs offer neither: each compiled graph fixes its query length, and the runtime owns the physical KV layout. We propose a tree verifier for this setting that packs repair rows and tree rows into one fixed 32-row call. Per-token rotary tables and an ancestor mask express the tree, and an exact-watermark rule re-scores accepted tokens whose KV is out of canonical order inside the next call, so no separate cache-movement call is needed. On the Rebellions ATOM+ NPU with the DFlash drafter, removing chunk-boundary replay first raises the linear baseline from 59.3 to 92.3~tokens/s. In a pre-specified study on 120 GSM8K questions, the tree verifier with a grouped 32-row projection head beats the fastest linear baseline by 12.2\% on Qwen3-4B (95\% CI 10.7--13.8\%) and by 13.2\% on Qwen3-8B (CI 12.1--14.4\%), and it cuts board energy per token by 10.0\% and 11.1\%. Changing the compiled verification shape also changes floating-point outputs on 55 of 120 questions for 4B and 9 for 8B, and a matched-prefix diagnostic traces the one lost 4B answer to the query-shape change alone.
\end{abstract}

\section{Introduction}

Speculative decoding trades parallel target-model work for fewer sequential decoding steps~\cite{leviathan2023fast,chen2023accelerating}. Under greedy decoding, a lightweight drafter proposes candidate tokens that the target model scores in a single pass, and the decoder keeps the prefix that matches the target's argmax choices. Tree-based variants pack multiple candidate branches into one verification call under an attention mask that exposes the committed prefix, each node itself, and its ancestors~\cite{miao2024specinfer,cai2024medusa,li2024eagle,chen2024sequoia}.

Ahead-of-time compilation changes the cost model of tree verification. On the Rebellions ATOM+ NPU, each compiled graph fixes its query length, so 16-row and 32-row verifiers are distinct binaries with different latencies, and the runtime owns the physical KV layout. A tree verifier on such a stack has to express tree attention and keep canonical KV state inside a fixed-shape graph. Drafting, vocabulary projection, and host-side selection must not erase the savings in target calls. Any change of compiled shape also changes decoded outputs, and that effect has to be measured.

Our key insight is that a fixed-shape call has spare rows: on ATOM+, a 32-row target call costs only 3.5\% more than a 16-row call. We use those rows to verify side branches of a draft tree and, in the same call, to repair accepted tokens whose KV landed out of order. We study this design with DFlash~\cite{dflash2026} as the drafter, Qwen3-4B and Qwen3-8B~\cite{qwen3} as targets, and GSM8K~\cite{cobbe2021gsm8k} as the workload. Our contributions are as follows.

\begin{enumerate}
\item \textbf{Cost observations on a static-graph NPU.} Removing chunk-boundary replay from a 16-row linear verifier, with the same compiled graph and identical outputs, raises throughput by 55.6\% (Section~\ref{sec:baseline}). Tree gains measured against replaying baselines are therefore overstated. Target latency grows by only 3.5\% from 16 to 32 rows and rises steeply beyond (Section~\ref{sec:costs}).
\item \textbf{Tree verification with in-call KV repair.} We propose a verifier that packs repair rows and tree rows into one 32-row call through per-token rotary tables, an ancestor mask, and an exact-watermark rule, together with a grouped 32-row projection head. Positive and negative controls check its correctness (Sections~\ref{sec:contracts} and~\ref{sec:costs}).
\item \textbf{Pre-specified confirmation and numerical diagnostics.} Under a protocol frozen before execution, the tree verifier beats the fastest tested linear baseline by 12.2\% on 4B and 13.2\% on 8B and cuts board energy per token by 10--11\%. We also trace where the accompanying output differences arise (Sections~\ref{sec:results} and~\ref{sec:numerics}).
\end{enumerate}

Rebellions has independently merged linear DFlash support for newer SDK releases~\cite{rbln975,rbln1020}, and our tree builder adapts the DDTree product-probability heap~\cite{ddtree2026}. This paper presents an end-to-end evaluation of tree verification on a static-graph NPU, with controls that separate tree branching from baseline design and compiled call shape.

\section{Background}
\label{sec:background}

\noindent\textbf{DFlash.} DFlash~\cite{dflash2026} uses a small block-diffusion model conditioned on target hidden states to predict a block of future tokens in parallel. With the public block-size-16 checkpoints for Qwen3-4B and Qwen3-8B, each linear verification block consists of one known anchor token followed by 15 draft predictions scored under greedy decoding.

\smallskip\noindent\textbf{Hardware and SDK.} We run on Rebellions ATOM+ (RBLN-CA22) PCIe cards under SDK 0.10.2, which compiles models into static graphs parameterized by a fixed query length. For rough roofline estimates in Section~\ref{sec:costs}, we use the advertised per-chip figures of 32~TFLOPS FP16 and 256~GB/s memory bandwidth from the ATOM product brief~\cite{atombrief}. Our Qwen3-4B experiments occupy a single card; Qwen3-8B places the target across two tensor-parallel cards and hosts the drafter and projection heads on a third.

\smallskip\noindent\textbf{Linear verification on a fixed-shape graph.} Let $L$ denote the committed sequence length and $q$ the compiled query length of a verification graph. A block-size-16 linear step fits directly into $q=16$, with valid KV entries occupying $[0, L)$. Our initial port instead aligned every call to a 64-token boundary: before scoring new draft tokens, it re-evaluated the $L \bmod 64$ committed tokens since the last multiple of 64 and split any step that crossed the next boundary. That alignment guaranteed chunk-aligned cache writes at the expense of extra target calls, a cost we quantify in Section~\ref{sec:baseline}.

\smallskip\noindent\textbf{Tree verification.} Tree verification packs $n$ candidate nodes into a single target call. Node $i$ at tree depth $d_i$ attends to the committed prefix, itself, and its ancestors; it therefore takes rotary position $L+d_i$ even though the runtime writes its KV entry to physical slot $L+i$. Once the target scores the tree, the decoder commits the longest accepted root-to-leaf path. Because our chosen NPU attention path writes consecutive KV slots for all $q$ query rows, rejected branches leave stale entries in the cache and accepted branches beyond a mismatch land out of canonical order. The caller must mask out rejected slots and restore accepted tokens to canonical positions.

\section{Direct-Start Linear Baseline}
\label{sec:baseline}

Before evaluating tree verification, we tested how much of our linear verifier's latency came from chunk-boundary replay. Table~\ref{tab:direct} compares three linear verifiers on 60 GSM8K prompts (one fresh process per arm, greedy decoding, up to 2048 generated tokens).

\begin{table}[t]
\centering
\caption{Linear verification with and without chunk-boundary replay (60 prompts, one process per configuration). R16 and D16 run the same compiled graph and produce identical tokens on all 60 prompts.}
\label{tab:direct}
\footnotesize
\setlength{\tabcolsep}{3pt}
\begin{tabular}{lrrrr}
\toprule
Configuration & tok/s & Time (s) & Correct & Target calls \\
\midrule
L64: $q$=64, replay & 64.85 & 304.7 & 58/60 & 4563 \\
R16: $q$=16, replay & 59.31 & 318.8 & 59/60 & 6549 \\
D16: $q$=16, direct & \textbf{92.29} & 204.9 & 59/60 & 3556 \\
\bottomrule
\end{tabular}
\end{table}

R16 and D16 execute the exact same 16-row graph and emit identical token sequences across all 60 prompts. L64 replays from 64-token boundaries and R16 from $16\lfloor L/16\rfloor$, while D16 starts verification directly at the logical cursor $L$. Starting at $L$ removes 2{,}993 redundant target calls and cuts cumulative target runtime from 246.1~s to 133.6~s. This accounts for 98.7\% of the 113.9~s wall-clock reduction and raises throughput by 55.6\%.

Chunk alignment proved unnecessary on this execution path. As long as the preceding KV prefix is valid, the compiled masked-prefill graph accepts unaligned starting positions within the single 4096-token KV block used by all evaluated contexts. All baselines and tree runs in this paper build on this fixed-shape masked-prefill operator. We did not use the SDK's multi-query decode operator, because our wrapped version of it produced incorrect hidden states on its first call at $q=4$ and $q=16$.

L64 cannot serve as a clean replay control for D16 because it changes both replay policy and compiled query shape ($q=64$ vs.\ $q=16$). The shape change alone alters the decoding trajectory: only 25 of 60 outputs are identical, so the 42\% throughput gap between L64 and D16 mixes the two effects. Isolating replay requires comparing R16 against D16. We therefore adopt direct-start verification (D16) as our baseline throughout the rest of the paper and discard preliminary tree numbers gathered against replaying verifiers.

\section{Verification Contracts}
\label{sec:contracts}

\begin{figure*}[t]
\centering
\includegraphics{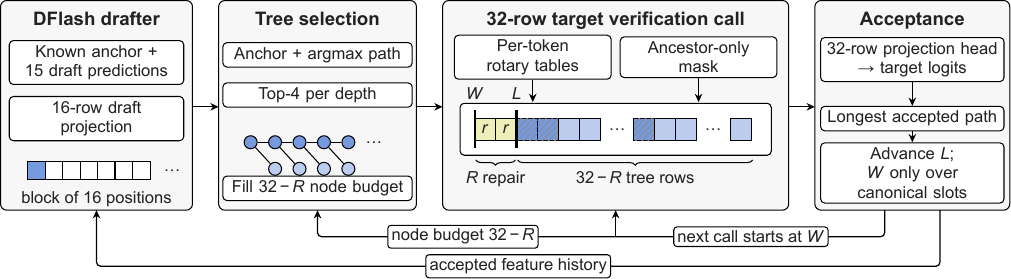}
\caption{One T32-H32 decoding step. The tree selector places the anchor and the full argmax path before high-probability branches, and one 32-row target call scores $R=L-W$ repair rows ahead of $32-R$ tree rows. The grouped 32-row projection head then extends $L$ and advances $W$ only over canonical slots (hatched cells: accepted path). Repair and tree verification share one fixed-shape call, so no separate cache-movement call is needed.}
\label{fig:system-overview}
\end{figure*}

\begin{figure}[t]
\centering
\includegraphics{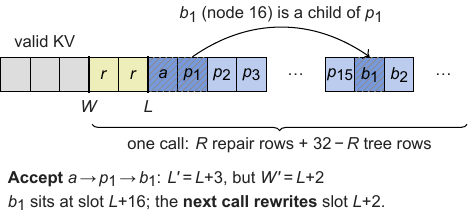}
\caption{Repair example with $R=2$: gray slots hold valid KV, yellow rows re-score committed tokens, and blue rows are tree candidates (hatched if accepted). As $b_1$ sits at slot $L+16$, $W$ stops at $L+2$ and the next call rewrites that slot.}
\label{fig:repair}
\end{figure}

Figure~\ref{fig:system-overview} illustrates the complete T32-H32 decoding loop. Every verification step runs a single compiled graph with $q=32$ query rows partitioned into repair rows and candidate tree rows (Figure~\ref{fig:repair}).

\smallskip\noindent\textbf{Repair rows.} Let the watermark $W$ be the exclusive upper bound of leading KV slots $[0, W)$ that already hold valid entries in canonical sequence order. When an accepted tree branch jumps over rejected sibling rows in the physical buffer, the newly committed length $L$ advances past $W$. The committed suffix $[W, L)$ then sits at non-canonical physical slots, while rejected nodes occupy the canonical slots in $[W, L)$. Instead of a separate cache-movement call, the next verification call begins at slot $W$ and places the $R = L - W$ unaligned committed tokens at the front of its query batch. Re-scoring those tokens in canonical order overwrites $[W, L)$ with valid KV entries inside the single mapped KV block.

\smallskip\noindent\textbf{Tree rows.} The remaining $32 - R$ rows carry speculative tree candidates. Following a selection rule chosen in an earlier GPU screen over fixed prefixes of 12 GSM8K prompts, the builder first inserts the known anchor token and all 15 sequential argmax predictions so the full greedy path is always present. It then fills the remaining budget from a max-heap keyed by cumulative factorized draft log-probabilities (from one DFlash pass, without branch re-conditioning), considering the top four tokens at each depth. Unlike the fixed-prefix GPU screen, the NPU decoder generates complete responses whose available tree budget $32-R$ varies step by step with the repair count. Once the target scores all 32 rows, the watermark advances only across the contiguous prefix of the accepted path that already landed in canonical slots: $W' = L + k$ for a contiguous canonical prefix of length $k$. Accepted tokens deeper on a side branch become repair rows on the next step.

Across 2{,}953 verification steps in a 60-prompt run, $R$ averaged 0.93 rows, so repair used few of the 32 rows and never needed a standalone pass. All contexts remained within a single 4096-token KV block.

Supporting this layout on SDK 0.10.2 required changes at the operator level:

\begin{itemize}
\item Each node must attend only to the committed prefix (including repair rows), itself, and its ancestors. Keeping \texttt{use\_position\_ids=True} in the inspected 0.10.2 wrapper routes execution to a strictly causal attention kernel, so passing a shallow configuration copy with \texttt{use\_position\_ids=False} only to the attention operator selects the explicit-mask \texttt{paged\_attn\_prefill} path.
\item Position IDs alone do not assign the depth-based rotary positions $L+d_i$ to tree nodes, and position lookup is bypassed on that attention path. The wrapper therefore supplies precomputed per-token cosine and sine tables directly while keeping the enclosing model configuration intact.
\item In an isolated probe of native cache updates, discarding the operator's opaque return tensor let the compiler pass from Torch export to internal IR remove the cache write as dead code. Keeping that opaque output live preserved the update, so our integrated verifier needs no separate commit call.
\end{itemize}

On test fixtures for both 4B and 8B, hidden states at the end of the repair region and all subsequent continuation tokens matched same-graph causal reconstruction bitwise. Shifting the repair boundary on purpose changed the continuation, while corrupting padding or off-path tree slots left valid outputs untouched. These fixtures confirm the repair and masking contracts, but a stricter cross-layout check did not pass: hidden states of tree rows differed from linear scoring of the same token sequence by up to 0.41 on 4B and 0.375 on 8B in maximum absolute error, above our preset equivalence threshold (Section~\ref{sec:numerics}).

\section{Execution Costs}
\label{sec:costs}

\subsection{Target verification latency by query shape}

\begin{figure}[t]
\centering
\includegraphics[width=\columnwidth]{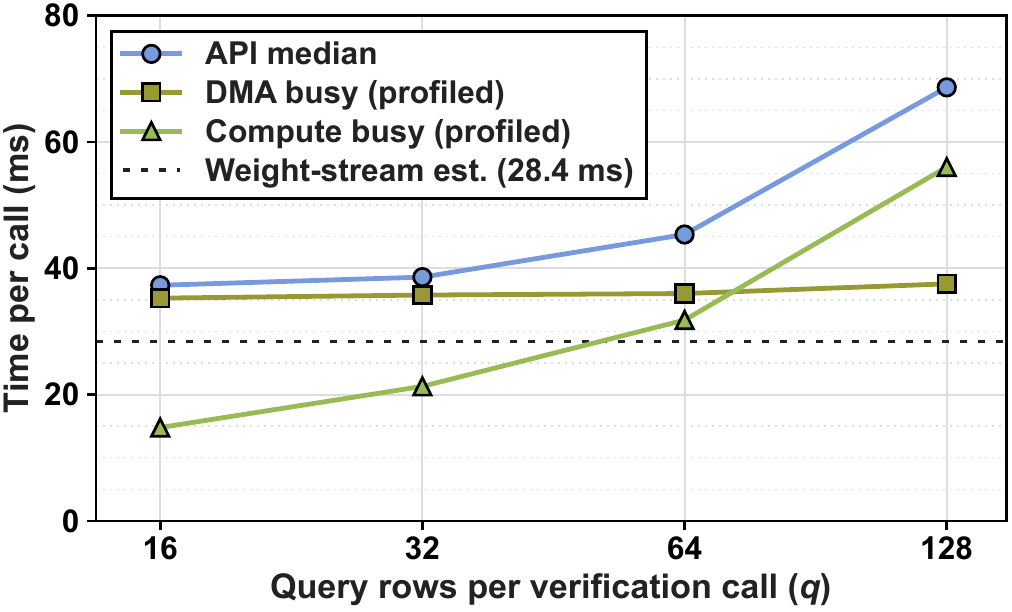}
\caption{4B target latency per call by query rows (prefix 256; API median of 30 calls; DMA and compute busy time from profiled traces). DMA time stays flat, so 32 rows cost only 3.5\% more than 16; beyond that, compute dominates.}
\label{fig:cost}
\end{figure}

Figure~\ref{fig:cost} plots the runtime API latency of a single 4B masked-prefill call across four compiled query shapes ($q \in \{16, 32, 64, 128\}$) at prefix length 256 with all rows active. We take medians over 30 unprofiled calls after a 12~s warm-up, excluding host input preparation. Separate profiled runs decompose device activity into the union of busy DMA and compute intervals.

Doubling the query batch from 16 to 32 rows adds only 1.3~ms (3.5\%). Going from 32 to 128 rows adds 30~ms. Profiled DMA activity stays nearly flat around 35--38~ms across all shapes, and compute time scales roughly linearly with $q$. A simple roofline estimate matches this crossover. Streaming the 3.63~billion non-embedding parameters once at two bytes per weight over 256~GB/s takes 28.4~ms, and dense matmul at 32~TFLOPS rises from 3.6~ms at $q=16$ to 29.1~ms at $q=128$.

This flat cost step between $q=16$ and $q=32$ motivates a 32-row verification tree. Extra target rows still have indirect costs, in output drift (Section~\ref{sec:numerics}) and in non-target work per step.

\subsection{Projection and host-side overheads}

Our initial 32-row tree verifier reused the baseline's 16-row vocabulary projection head. On the 60-prompt set, it beat linear verification inside the same 32-row graph by 8.6\% (88.3 vs.\ 81.4~tok/s) yet trailed the 16-row direct-start baseline (92.3~tok/s) by 4.3\%. Average accepted tokens per step (including the anchor) rose from 5.45 to 6.70, and cumulative target time dropped from 149.1~s to 123.5~s. However, invoking the 16-row projection head twice per 32-row step pushed projection time from 33.2~s to 41.9~s, and host tree selection grew from 6.3~s to 11.4~s.

To remove the split projection bottleneck, we compiled a 32-row projection head and tested both head sizes across linear and tree verifiers in a separate 60-prompt exploratory sweep (one process per arm). In this sweep, switching the 32-row tree to the 32-row head lifted its throughput from 88.5~tok/s (the configuration above, rerun) to 92.4~tok/s. Total head calls fell from 8{,}919 to 5{,}912, because draft and prompt-bonus projections still used 16-row calls. Head runtime-API time fell from 37.4~s to 26.0~s, and end-to-end projection time including tensor preparation fell from 42.1~s to 30.8~s. On linear verification, the 32-row head shifted single-run throughput by $+1.3\%$ at $q=16$ (reaching 93.5~tok/s, still 1.2\% ahead of the tree's 92.4~tok/s) and by $-2.0\%$ at $q=32$, within the noise of single-process runs.

Two other head variants explored earlier offered no benefit and were dropped before confirmation. A 64-row head in a six-prompt replay prototype altered two outputs and gave virtually identical runtime on the remaining four matched prompts (44.59~s vs.\ 44.61~s). A hybrid head that offloaded part of the argmax reduction to the host matched all 144 tested rows but raised median API latency from 4.76~ms to 13.24~ms.

\section{Confirmation}
\label{sec:results}

In the single-process exploratory sweep above, aggregate throughput still slightly favored linear verification (93.5 vs.\ 92.4~tok/s). Per-prompt inspection showed that one outlier drove that result. On one question, the tree's altered trajectory failed to terminate early and ran to the 2048-token cap at 43~tok/s, consuming 22.5\% of the tree arm's total runtime; linear verification finished it in 874 tokens. Across the other 59 questions the tree achieved 106.7~tok/s against 94.0~tok/s, and across all 60 questions the geometric mean of paired per-question throughput ratios was 1.12. Filtering out that outlier after observing it would have been post hoc, so we locked a confirmation protocol before running it on a fresh question sample.

\smallskip\noindent\textbf{Protocol.} Using seed 20261013, we sampled 120 questions uniformly without replacement from the 819 GSM8K test items untouched by earlier exploratory or warm-up runs. We compared the 32-row tree with the 32-row head (T32-H32) against two direct-start 16-row linear controls using either head size (D16-H16 and D16-H32), keeping frozen FP16 artifacts for the drafter and heads. Each arm ran three independent processes ordered by a Latin square, with a 20~s warm-up and identical per-repeat question ordering across arms. Requests ran serially under greedy decoding (thinking disabled, 2048-token cap), and all requests succeeded without hitting the token limit. Timed generation spans prompt prefill, drafting, target verification, projection, and host selection. It excludes model loading, tokenization, warm-up, and answer grading. Qwen3-8B then ran on the same 120-question list to test transfer across model scale and multi-card placement.

We pre-declared two co-primary metrics: aggregate throughput (total generated tokens divided by total generation time) and the geometric mean of paired per-question throughput ratios. Nonparametric 95\% confidence intervals use 10{,}000 bootstrap resamples drawn at the whole-question level, carrying all three process repeats together. A speedup claim required both lower bounds to exceed 1.0 against both linear controls. During generation, we sampled gross board power every 0.25~s on the occupied NPU cards and integrated it into energy.

\begin{table*}[t]
\centering
\caption{Confirmation runs (120 questions $\times$ 3 process repeats per arm; time and tokens summed over repeats). Energy is gross board energy over the occupied cards (one for 4B, three for 8B). Changed outputs count prompts whose repeat-1 output differs from D16-H16.}
\label{tab:confirm}
\small
\setlength{\tabcolsep}{5pt}
\begin{tabular}{llrrrrrr}
\toprule
Model (placement) & Arm & tok/s & Time (s) & Tokens & Correct (each repeat) & J/token & Changed outputs \\
\midrule
Qwen3-4B (1 card) & D16-H16 & 94.71 & 1173.1 & 111{,}108 & 104 & 0.590 & -- \\
 & D16-H32 & 93.77 & 1181.7 & 110{,}808 & 104 & 0.593 & 20/120 \\
 & T32-H32 & \textbf{106.24} & 1046.3 & 111{,}156 & 103 & \textbf{0.531} & 55/120 \\
\midrule
Qwen3-8B (TP2 + 1) & D16-H16 & 79.11 & 1433.4 & 113{,}391 & 104 & 1.694 & -- \\
 & D16-H32 & 78.22 & 1449.7 & 113{,}391 & 104 & 1.713 & 0/120 \\
 & T32-H32 & \textbf{89.56} & 1269.2 & 113{,}673 & 105 & \textbf{1.507} & 9/120 \\
\bottomrule
\end{tabular}
\end{table*}

\begin{figure}[t]
\centering
\includegraphics[width=\columnwidth]{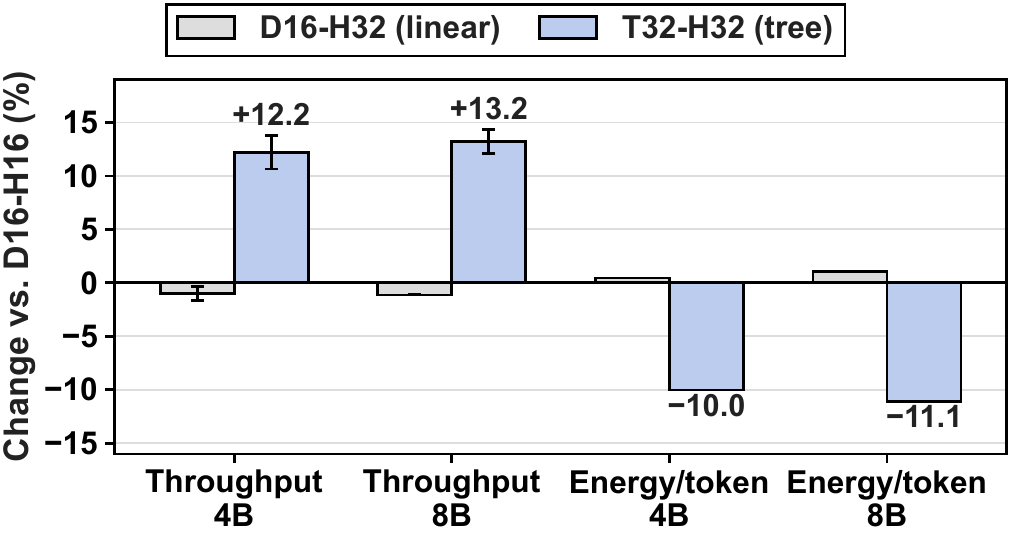}
\caption{Change vs.\ D16-H16 (aggregate throughput with 95\% bootstrap CI; energy: gross board J/token). The tree gains 12--13\% throughput and saves 10--11\% energy on both models; the 32-row head alone does neither.}
\label{fig:confirm}
\end{figure}

Table~\ref{tab:confirm} and Figure~\ref{fig:confirm} summarize the confirmation runs. On Qwen3-4B, T32-H32 beats the faster linear control (D16-H16) by 12.2\% in aggregate throughput (95\% CI 10.7--13.8\%) and by 12.1\% in per-question geometric mean (CI 11.0--13.4\%). Against D16-H32 the gains are 13.3\% and 13.2\%. On Qwen3-8B, T32-H32 leads D16-H16 by 13.2\% in aggregate (CI 12.1--14.4\%) and 13.3\% in geometric mean (CI 12.2--14.6\%), and it leads D16-H32 by 14.5\%. Every confidence interval clears unity on both models, which satisfies our pre-declared criterion. Gross board energy per token drops by 10.0\% on 4B (0.590 to 0.531~J) and by 11.1\% on 8B (1.694 to 1.507~J). Within every arm, all three process repeats produced identical token sequences with valid power logs throughout.

T32-H32 changes output text on 55 of 120 prompts for 4B and 9 of 120 for 8B, so these metrics compare two end-to-end decoders, not two bit-exact implementations of the same mapping. Swapping only the projection head (D16-H32) already changes 20 of 120 outputs on 4B, although none on 8B. Scored GSM8K accuracy shifted by a single question in opposite directions: from 104 to 103 on 4B (one lost, none gained) and from 104 to 105 on 8B (one gained, none lost). The exact McNemar test gives $p=1$ in both cases.

\section{Numerical Behavior}
\label{sec:numerics}

Under greedy argmax decoding, tiny floating-point perturbations flip the chosen token whenever the top two logits are nearly tied. The diagnostics below ask whether this drift needs a tree at all, locate where it first appears inside a layer, and trace the single lost answer on 4B.

Query shape alone alters hidden states, even for linear inputs. Scoring identical linear token paths from fixed prefixes across $q \in \{16, 32, 64\}$ shifts final-layer hidden states by up to 0.81 in maximum absolute value (about 2\% in relative $L_2$). Switching between tree and linear layouts at the same compiled $q$ shifts them by up to 0.50 (about 1\%). Neither perturbation flipped an argmax across the five short test paths per prefix. The linear spread alone already exceeds the strict tolerance that failed in Section~\ref{sec:contracts}.

The divergence starts in attention. We compiled the first transformer layer of Qwen3-4B with intermediate tensor taps after each sub-operation. RMSNorm outputs, raw Q/K/V projections, and rotary-embedded Q/K were bitwise identical between tree and linear layouts of the same tokens. The first discrepancy appeared at the attention output ($2.4\times10^{-4}$ max absolute difference) and grew through the output projection and MLP. The taps slightly perturb the compiled layer itself ($3.9$--$7.8\times10^{-3}$ at the layer output), so this probe localizes the split in the instrumented binary. A changed reduction order inside attention is the likely cause.

The lost 4B answer traces back to query shape. On the single confirmation prompt that T32-H32 missed and both linear controls solved, all three arms generated the same 269 tokens before diverging. At the next step the top-two logit margin was 0.0625 in each arm, with opposite winners. Scoring that 269-token prefix linearly through the 32-row target graph reproduced T32-H32's hidden state bitwise and selected the same diverging token. Switching the compiled target graph from 16 to 32 rows is thus enough to trigger this branch, without any contribution from tree masking or non-canonical KV slots.

\section{Related Work}
\label{sec:related}

Speculative sampling~\cite{leviathan2023fast,chen2023accelerating} was generalized to ancestor-masked token trees by SpecInfer~\cite{miao2024specinfer}, Medusa~\cite{cai2024medusa}, and EAGLE~\cite{li2024eagle}. Sequoia~\cite{chen2024sequoia} and BASTION~\cite{bastion2026} shape draft trees around hardware verification latency, a principle we follow in choosing the $q=32$ cost step. For block-diffusion drafters, DDTree~\cite{ddtree2026} ranks candidate prefixes by factorized single-pass probabilities. We adopt the same heap objective but limit expansion to the top four tokens per depth and seed the full greedy path, so DDTree's optimality proof for unconstrained trees no longer applies.

Static tree attention has also been shown on NPUs. Dai et al.~\cite{pangu2026npu} port Medusa to OpenPangu-7B using precomputed tree topologies, retrieval tensors, and on-device path gathering, and report a $1.35\times$ speedup on short generations. Instead of a fixed Medusa topology, we pair a block-diffusion drafter with per-step tree construction and same-call watermark repair on the ATOM+ runtime, measured against direct-start linear baselines. On GPUs, DScale~\cite{dscale2026} reduces DFlash padding overhead from the other side, shrinking verification work through path-aware tiles and adaptive lengths. We instead use spare static-graph rows to verify side branches under a KV-repair contract.

Rebellions merged native DFlash support into \texttt{vllm-rbln}~\cite{rbln975,rbln1020} for SDK releases newer than 0.10.2. That integration projects only newly accepted context without draft-history replay, and it includes a configuration check that rejects causal drafters. The check concerns the draft-model architecture, not target-side tree verification. Our study was conducted independently on SDK 0.10.2.

\section{Limitations}

Our findings are scoped to a single accelerator family, SDK 0.10.2, two Qwen3 model sizes, and greedy GSM8K decoding. Compiling 16- and 32-row target verification graphs required bypassing a high-level shape check inside our standalone research wrapper, so these shapes are not supported out-of-the-box configurations. The Qwen3-8B experiment changes model scale and multi-card placement together and reuses the 120-question sample from 4B, and 120 questions lack the statistical power to rule out small accuracy regressions. The bootstrap intervals capture question sampling, not machine-to-machine variance. Energy figures integrate gross board power on occupied NPU cards without subtracting idle draw or adding host CPU/DRAM power.

Several analyses are partial. Correctness controls compare model outputs and do not read back raw device KV buffers. The cumulative timers in Section~\ref{sec:costs} come from full runs whose trajectories differ, so they do not isolate fixed-work operator speedups, and we did not collect memory-traffic counters to confirm the projection-head explanation. The roofline estimates assume FP16 at advertised peak rates, while the host passes FP32 tensors and we did not inspect compiled kernel precision or utilization. On the numerical side, we neither inspected attention kernel internals nor repeated the matched-prefix trace across the remaining changed trajectories. We also left out three engineering comparisons: KV compaction via the runtime's explicit copy primitive, the OpenPangu NPU stack, and the newer vendor plugin. Finally, the end-to-end runs evaluate rotary tables, ancestor masking, in-call repair, and the grouped head as one bundle, without isolating each component's contribution.

\section{Conclusion}

On a commercial static-graph NPU, 32-row tree verification with in-call KV repair improved DFlash decoding throughput by 12--13\% and lowered board energy per token by 10--11\% relative to our fastest direct-start linear baseline. That baseline was itself 55.6\% faster than the replaying verifier we started from, so the choice of baseline matters as much as the tree. The net speedup came from explicit rotary tables, ancestor masking, and same-call watermark repair together with a grouped 32-row projection head. Compiled graph shapes also shift floating-point outputs and caused at least one changed answer on their own. A speculative decoder on static-graph hardware should therefore report its verification shape along with its drafter and tree.

\section*{Acknowledgments}
This research was supported by the Ministry of Science and ICT (MSIT) through the 2026 High-Performance Computing (HPC) Support Program (RQT-26-041563).

\bibliographystyle{plain}
\bibliography{refs}

\begin{thebibliography}{10}

\bibitem{cai2024medusa}
Tianle Cai, Yuhong Li, Zhengyang Geng, Hongwu Peng, Jason~D. Lee, Deming Chen,
  and Tri Dao.
\newblock Medusa: Simple {LLM} inference acceleration framework with multiple
  decoding heads.
\newblock In {\em Proceedings of the 41st International Conference on Machine
  Learning (ICML)}, 2024.

\bibitem{chen2023accelerating}
Charlie Chen, Sebastian Borgeaud, Geoffrey Irving, Jean-Baptiste Lespiau,
  Laurent Sifre, and John Jumper.
\newblock Accelerating large language model decoding with speculative sampling.
\newblock {\em arXiv preprint arXiv:2302.01318}, 2023.

\bibitem{dflash2026}
Jian Chen, Yesheng Liang, and Zhijian Liu.
\newblock {DFlash}: Block diffusion for flash speculative decoding.
\newblock {\em arXiv preprint arXiv:2602.06036}, 2026.

\bibitem{dscale2026}
Rongjian Chen, Minxian Xu, Zhengxin Fang, Kejiang Ye, and Chengzhong Xu.
\newblock {DScale}: Scaling block-diffusion speculative decoding with adaptive
  verification.
\newblock {\em arXiv preprint arXiv:2609.37532}, 2026.

\bibitem{chen2024sequoia}
Zhuoming Chen, Avner May, Ruslan Svirschevski, Yuhsun Huang, Max Ryabinin,
  Zhihao Jia, and Beidi Chen.
\newblock Sequoia: Scalable and robust speculative decoding.
\newblock {\em arXiv preprint arXiv:2402.12374}, 2024.

\bibitem{cobbe2021gsm8k}
Karl Cobbe, Vineet Kosaraju, Mohammad Bavarian, Mark Chen, Heewoo Jun, Lukasz
  Kaiser, Matthias Plappert, Jerry Tworek, Jacob Hilton, Reiichiro Nakano,
  Christopher Hesse, and John Schulman.
\newblock Training verifiers to solve math word problems.
\newblock {\em arXiv preprint arXiv:2110.14168}, 2021.

\bibitem{pangu2026npu}
Yuntao Dai, Jing Wu, Hang Gu, and Teng Wang.
\newblock Accelerating {OpenPangu} inference on {NPU} via speculative decoding.
\newblock {\em arXiv preprint arXiv:2603.03383}, 2026.

\bibitem{leviathan2023fast}
Yaniv Leviathan, Matan Kalman, and Yossi Matias.
\newblock Fast inference from transformers via speculative decoding.
\newblock In {\em Proceedings of the 40th International Conference on Machine
  Learning (ICML)}, 2023.

\bibitem{li2024eagle}
Yuhui Li, Fangyun Wei, Chao Zhang, and Hongyang Zhang.
\newblock {EAGLE}: Speculative sampling requires rethinking feature
  uncertainty.
\newblock In {\em Proceedings of the 41st International Conference on Machine
  Learning (ICML)}, 2024.

\bibitem{miao2024specinfer}
Xupeng Miao, Gabriele Oliaro, Zhihao Zhang, Xinhao Cheng, Zeyu Wang, Zhengxin
  Zhang, Rae Ying~Yee Wong, Alan Zhu, Lijie Yang, Xiaoxiang Shi, Chunan Shi,
  Zhuoming Chen, Daiyaan Arfeen, Reyna Abhyankar, and Zhihao Jia.
\newblock {SpecInfer}: Accelerating large language model serving with
  tree-based speculative inference and verification.
\newblock In {\em Proceedings of the 29th ACM International Conference on
  Architectural Support for Programming Languages and Operating Systems
  (ASPLOS)}, 2024.

\bibitem{bastion2026}
Soowon Oh, Nam Cao, Yujin Kim, Hojung Jung, Huzama Ahmad, Sangmin Bae, and
  Se-Young Yun.
\newblock {BASTION}: Budget-aware speculative decoding with tree-structured
  block diffusion drafting.
\newblock {\em arXiv preprint arXiv:2605.29727}, 2026.

\bibitem{atombrief}
{Rebellions}.
\newblock {ATOM} product brief, v3.1.
\newblock
  \url{https://rebellions.ai/wp-content/uploads/2023/09/Rebellions_ATOMProduct_Brief_v.3.1.pdf},
  2023.

\bibitem{rbln1020}
{Rebellions}.
\newblock vllm-rbln pull request \#1020: feature(specdec): enable {DFlash}
  speculative decoding with max\_num\_seqs $>$ 1.
\newblock \url{https://github.com/RBLN-SW/vllm-rbln/pull/1020}, 2026.
\newblock Merged 2026-09-21.

\bibitem{rbln975}
{Rebellions}.
\newblock vllm-rbln pull request \#975: feature(specdec): {DFlash} parallel
  drafting on {RBLN}.
\newblock \url{https://github.com/RBLN-SW/vllm-rbln/pull/975}, 2026.
\newblock Merged 2026-09-03.

\bibitem{ddtree2026}
Liran Ringel and Yaniv Romano.
\newblock Accelerating speculative decoding with block diffusion draft trees.
\newblock {\em arXiv preprint arXiv:2604.12989}, 2026.

\bibitem{qwen3}
An~Yang et~al.
\newblock {Qwen3} technical report.
\newblock {\em arXiv preprint arXiv:2505.09388}, 2025.

\end{thebibliography}

\end{document}